\documentclass[aps,prd,amsfonts,amssymb,preprint,eqsecnum,nofootinbib,superscriptaddress]
{revtex4} 
\usepackage{graphics,epsfig}

\newcommand{\pslash}{p\llap{/\kern-0.3pt}}
\newcommand{\qslash}{q\llap{/\kern-0.3pt}}
\newcommand{\rslash}{r\llap{/\kern-0.3pt}}
\newcommand{\lslash}{\ell\llap{/\kern-0.3pt}}

\newcommand{\Tr}{{\rm Tr}}

\begin{document}
\preprint{WM-07-111}
%
\title{\vspace*{0.5in}
An Exceptional Electroweak Model
\vskip 0.1in}
\author{Christopher D. Carone}\email[]{cdcaro@wm.edu}
\author{Ashwin Rastogi}\email[]{axrast@wm.edu}
\affiliation{Particle Theory Group, Department of Physics,
College of William and Mary, Williamsburg, VA 23187-8795}
\date{December 2007}
\begin{abstract}
We consider a gauge extension of the electroweak sector of the Standard Model based on the 
group G$_2 \times$SU(2)$\times$U(1). The exceptional group G$_2$ is the smallest rank two group 
that contains SU(3) as a subgroup; the SU(3) prediction $\sin^2\theta_w=1/4$ follows approximately 
in this model if the couplings of the additional SU(2) and U(1) factors are sufficiently large.  We 
study the symmetry breaking sector of the model, the bounds from precision electroweak constraints 
and the mass spectrum of exotic gauge bosons that may be produced at future colliders.  We also discuss
an SU(3) electroweak model in which a vector-like sector is included explicitly to facilitate the 
decays of otherwise stable exotic states. The models considered here represent plausible extensions 
of the minimal SU(3) electroweak model with potentially distinctive TeV-scale phenomenology.
\end{abstract}
\maketitle

\section{Introduction} \label{sec:intro}
A major paradigm in the study of physics beyond the Standard Model is the assumption
that the forces of nature should have a simple, unified description at high 
energies.  Four-dimensional Grand Unified Theories (GUTs) achieve this goal by embedding
the Standard Model gauge group into a larger group, such as SU(5) or SO(10).  Gauge invariance 
requires that the matter and Higgs fields of the Standard Model appear somewhere within 
complete representations of the unified group.  Perhaps the most compelling ``experimental'' 
evidence in favor of grand unification is the observation that the Standard Model gauge
couplings do indeed unify around $2\times 10^{16}$~GeV, at least if the low-energy particle 
content is that of the Minimal Supersymmetric Standard Model, and an appropriate normalization
of hypercharge is assumed.  It is not uncommon for practitioners in the field to remark that 
it would be surprising if this unification of coupling constants turns out to be purely 
accidental.

In the same spirit, it is intriguing that the SU(2) and U(1) gauge couplings in the miminal 
Standard Model, $g$ and $g'$, satisfy relation $g\approx\sqrt{3} \,g'$, or equivalently
\begin{equation}
\sin^2\theta_w \approx 1/4 \,\,\,,
\label{eq:sstp}
\end{equation}
around $4$~TeV.  The possibility that this might not be accidental has motivated the
exploration of models in which there is a partial unification of the electroweak gauge
groups at the TeV scale~\cite{weinberg,kd,cekt,kd2,su3w}.  Specifically, Eq.~(\ref{eq:sstp}) is 
reproduced if SU(2)$\times$U(1) is embedded in the group SU(3) and the normalization of hypercharge 
is chosen to be consistent with the branching rule
\begin{equation}
{\bf 3} = {\bf 2_{1/2}} + {\bf 1_{-1}} \,\,\,.
\label{eq:br3}
\end{equation}
The SU(2)$\times$U(1) representations on the right-hand-side of Eq.~(\ref{eq:br3}) allow for
the embedding of the left- and right-handed leptons of the Standard Model,
\begin{equation}
{\bf 3}_R \sim \left(\begin{array}{c} \hspace{0.5em}(e_L)^c \\ -(\nu_L)^c 
\\ \hspace{0.5em}e_R \end{array}\right)  \,\,\,,
\label{eq:lepem}
\end{equation}
where the superscript $c$ represents charge conjugation.  The immediate problem with SU(3) 
electroweak unification (aside from anomaly cancellation) is that there are no SU(3) 
representations that contain the quark fields of the Standard Model.  

One solution to this problem is to extend the gauge group to 
\begin{equation}
G = \mbox{SU(3)}\times\mbox{SU(2)}\times\mbox{U(1)},
\label{eq:su3grp}
\end{equation}
with gauge couplings $g_3$, $\tilde{g}$ and $\tilde{g}'$.  Let SU(2)$_0\times$U(1)$_0$ represent 
the subgroup of SU(3) defined in our previous example.   If the electroweak gauge group of 
the Standard Model is now identified as the diagonal subgroup of SU(2)$_0\times$U(1)$_0$ and 
the additional SU(2)$\times$U(1) factors in Eq.~(\ref{eq:su3grp}), then one finds that
\begin{eqnarray}
\frac{1}{g^2} &=& \frac{1}{g_3^2} + \frac{1}{\tilde{g}^2} \nonumber \\
\frac{1}{{g'}^2} &=& \frac{3}{g_3^2} + \frac{1}{{\tilde{g}}'^2} \,\,\,,
\label{eq:match}
\end{eqnarray}
where $g$ and $g'$ are the Standard Model SU(2)$_W$ and U(1)$_Y$ gauge couplings. The 
prediction that $\sin^2\theta_w \approx 1/4$ follows if $\tilde{g}$ and ${\tilde{g}}'$ are 
much larger than $g_3$.  Standard Model fields transform only under the additional 
SU(2)$\times$U(1) factors and are given their conventional charge assignments.  All Standard
Model fields can therefore be incorporated without anomalies.  Without embedding the leptons
into a ${\bf 3}$ of SU(3), the identification of hypercharge is fixed by the charge 
assignment\footnote{Note that ${\bf \bar{2}}$ is isomorphic to ${\bf 2}$.  See Eq.~(\ref{eq:ttb}).} 
of the Higgs field $\Sigma$ whose vacuum expectation value (vev) breaks 
$\mbox{SU(3)}\times\mbox{SU(2)}\times\mbox{U(1)} \rightarrow \mbox{SU(2)}_W \times \mbox{U(1)}_Y$:
\begin{equation}
\Sigma \sim ({\bf 3},{\bf {\bar{2}}_{-1/2}})  \,\,\,\,\, , \,\,\,\,\,
\langle \Sigma \rangle = \left(\begin{array}{cc} M & 0\\ 0 & M \\ 0 & 0 \end{array}\right)
\,\,\,.
\label{eq:sigdef}
\end{equation}
This SU(3) electroweak model, proposed originally by Kaplan and Dimopolous~\cite{kd}, represents 
a simple gauge extension of the Standard Model that provides an interesting physical prediction, 
Eq.~(\ref{eq:sstp}), over a wide range of the model's parameter space.  It is interesting
to note that the gauge structure in Eq.~(\ref{eq:su3grp}) follows from the two-site deconstruction 
of a five-dimensional (5D) SU(3) unified gauge theory with symmetry breaking imposed via boundary 
conditions.  The 4D SU(3) electroweak model was studied in phenomenological detail by 
Cs\'{a}ki {\em et al.}~\cite{cekt}, and motivated the study of 5D SU(3) electroweak unification
by a number of authors~\cite{kd2,su3w}. 

It is natural to consider other simple 4D models with the structure $G_U \times$SU(2)$\times$U(1)
that predict $\sin^2\theta_w \approx 1/4$.  Of course, a theory can be constructed
for any $G_U$ that contains SU(3) as a subgroup; the larger the group $G_U$, the more exotic
states one expects at the scale of symmetry breaking.   Here we focus on the next-to-minimal choice 
for the group $G_U$.  The only groups of rank $2$ that contain an SU(3) subgroup are the groups G$_2$ 
and SU(3) itself. Moreover, the exceptional group G$_2$ has $14$ generators, a number smaller than 
that of any Lie group of rank greater than $2$ that contains an SU(3) subgroup.  Motivated by these
observations, we present a G$_2\times$SU(2)$\times$U(1) model that predicts 
$\sin^2\theta_w \approx 1/4$, and we study the constraints on the parameter space of the 
model from precision electroweak observables.   The group $G_2$ has been used before in 
extensions of the electroweak sector of the Standard Model~\cite{manton,cgm}, most notably in 
the six-dimensional gauge-Higgs unification model of Ref.~\cite{cgm}.  By contrast, the model 
studied here corresponds to the deconstruction of a 5D unified G$_2$ model with gauge symmetry 
broken by (non-orbifold) boundary conditions.  Nonetheless, the analysis here will proceed from a 
purely four-dimensional perspective.

One issue that must be considered in models with the gauge group structure $G_U \times$SU(2)$\times$U(1) 
is the appearance of exotic, stable charged states~\cite{cekt}.  For example, the charged gauge bosons in 
$G_U$ that are not contained in the SU(2)$_0\times$U(1)$_0$ subgroup potentially fall in this category.  
New stable charged particles are severely constrained by heavy isotope searches~\cite{his} as well 
as cosmological bounds~\cite{cosmo}.  Enlarging $G_U$ from SU(3) to G$_2$ (or to any larger gauge group) only 
exacerbates the problem. We first address this issue in the context of an SU(3) model; we
introduce vector-like leptons into which exotic bosons may decay, and include new Higgs fields
that allow these vector-like states to mix with their Standard Model counterparts.  The inclusion of 
additional fields and vevs alters that analysis of unification and electroweak constraints described in 
Ref.~\cite{cekt}.  We determine the constraints on this alternative SU(3) electroweak model, taking into account 
updated electroweak data from LEP II that was not available at the time of Ref.~\cite{cekt}.   With this
machinery in place, we find the analogous constraints on the G$_2\times$SU(2)$\times$U(1) model.
In the SU(3) model, we will explicitly integrate out the vector-like sector to induce a set of higher-dimension
operators that allow otherwise stable states to decay.  In the G$_2$ model, we will construct a similar
set of operators directly.  Interestingly, both the SU(3) and G$_2$-based models have exotic, doubly-charged 
gauge bosons that may be long-lived; in principle, these states could travel a macroscopic distance before 
undergoing a lepton-flavor-violating decay.  This would be a remarkable experimental signature.  Unfortunately,
one cannot determine the lifetime without specifying parameters that are not determined in the
low-energy theory.  

Our paper is organized as follows.  In Section~II, we define the SU(3)$\times$SU(2)$\times$U(1)
and the G$_2\times$SU(2)$\times$U(1) models that we study in this paper.  In Section~III, 
we find the parameter space of these models that is allowed by precision electroweak constraints.  
In Section~IV we discuss the effective interactions that contribute to the decays of heavy gauge boson 
states. In Section~V, we summarize our conclusions.

\section{The Models}\label{sec:two}
\subsection{$G_U=$SU(3)} \label{sec:su3model}

In this section, we extend the matter and Higgs field content of the minimal SU(3) 
electroweak model~\cite{kd,cekt}.  Our symmetry-breaking sector consists of two 
fields, $\Sigma$ and $\chi$:  
\begin{equation}
 \Sigma \sim (\mathbf{3},\mathbf{\bar{2}_{-1/2}}), 
\quad \langle\Sigma\rangle = \left(\begin{array}{cc} M &0 \\ 0& M \\ 0 &0\end{array}\right)\,\,\,,
\label{eq:mdef1}
\end{equation}
\begin{equation}
 \chi \sim (\mathbf{3},\mathbf{1_{1}}), 
\quad \langle\chi\rangle = \left(\begin{array}{c} 0 \\ 0 \\ x\, M \end{array}\right)\,\,\,.
\label{eq:xdef1}
\end{equation}
The parameter $M$ sets the scale of symmetry breaking, while $x$ indicates the ratio of the $\chi$ 
and $\Sigma$ vevs.  The pattern of vevs shown breaks the gauge group down to the diagonal subgroup of 
SU(2)$_0\times $U(1)$_0$ and SU(2)$\times$U(1).  We identify the unbroken symmetry as the electroweak gauge 
group of the Standard Model, SU(2)$_W \times $U(1)$_Y$.  In addition, we include $n_F$ pairs of vector-like 
fermions 
\begin{equation}
\psi^a_L \sim \psi^a_R \sim {\bf 3},\,\,\,\,\,\,\,\,\, a=1\ldots n_F \,,
\end{equation}
with mass $M_F$.  The mass scale $M_F$ is a free parameter that we will assume is either of the same
order as $M$ or much heavier.  Any of the SU(3) gauge bosons can decay to pairs of the vector-like 
fermions, if the decay is kinematically allowed.  More importantly, each component of $\psi$ will mix
with a Standard Model lepton field as a consequence of the choice of Higgs field representations in 
Eqs.~(\ref{eq:mdef1}) and (\ref{eq:xdef1}).  The portion of the Lagrangian involving the vector-like 
fields is given by
\begin{equation}
{\cal L} = \overline{\psi} (i \not\!\!D - M_F) \psi - \left[\overline{\psi}_L \Sigma \lambda^\ell \ell_R
+ \overline{\psi}_L \chi \lambda^e e_R + \mbox{ h.c.}\right] \,\,\,,
\label{eq:hlmix}
\end{equation}
where $\ell_R$ and  $e_R$ refer to Standard Model SU(2)$_W$ doublet and singlet leptons, appropriately 
charge conjugated as in Eq.~(\ref{eq:lepem}); $\lambda^\ell$ and $\lambda^e$ are Yukawa matrices that 
mix the heavy and light states.  Eq~(\ref{eq:hlmix}) assures that the exotic gauge fields will decay 
ultimately to Standard Model leptons.   The simplest way to see this is to integrate out the vector-like
matter and study the resulting effective interactions.  This will be done in Section~\ref{sec:four}.  For 
now, we finish defining the model, treating $n_F$ and $M_F$ as free parameters.

The SU(3) gauge bosons $A=A^a T^a$ transform in the adjoint representation, which decomposes under 
$SU(2)\times U(1)$ as
\begin{equation}
\mathbf{8}=\mathbf{3_0}+\mathbf{1_0}+\mathbf{2_{3/2}}+\mathbf{2_{-3/2}}.
\end{equation}
The $\mathbf{2_{3/2}}$ and $\mathbf{2_{-3/2}}$ representations appear in the Lagrangian as a single complex 
doublet.  Taking into account the additional SU(2)$\times$U(1) factors, we deduce that the gauge bosons in 
the model fall in the following SU(2)$_W\times$U(1)$_Y$ representations: 
\begin{equation}
\mathbf{3_0}+\mathbf{1_0}+\mathbf{3_0}+\mathbf{1_0}+\mathbf{2_{\pm3/2}} \,\,\,.
\end{equation}
The mass eigenstate $\mathbf{3_0}$ fields arise from mixing of the SU(2) gauge fields $\widetilde{W}^a$ and
the SU(2)$_0$ gauge fields $A^a$, for $a=1,2,3$.  In the $(A^a,\widetilde{W}^a)$ basis, the mass squared 
matrix is
\begin{equation}
M^2 \left(\begin{array}{cc} g_3^2 & -g_3\tilde{g} \\ -g_3\tilde{g} & \tilde{g}^2 \end{array}\right).
\end{equation}
Therefore, one obtains the mass eigenstates
\begin{eqnarray}
W^a_L&=&c_\phi A^a - s_\phi \widetilde{W}^a \\
W^a_H&=&s_\phi A^a + c_\phi \widetilde{W}^a 
\end{eqnarray}
with
\begin{equation}
s_\phi=\frac{-g_3}{\sqrt{g_3^2+\tilde{g}^2}}\quad \mbox{ and } \quad
c_\phi=\frac{\tilde{g}}{\sqrt{g_3^2+\tilde{g}^2}} \,\,\,,
\end{equation}
and the masses 
\begin{eqnarray}
M_{W_L}&=&0 \\
M_{W_H}&=& (g_3^2+\tilde{g}^2)^{1/2}\,M.
\label{eq:whmass}
\end{eqnarray}
Note that our notation follows that of Ref.~\cite{cekt}.   The mass eigenstate $\mathbf{1_0}$ fields arise 
from the mixing of the U(1) field $\widetilde{B}$ and the U(1)$_0$ field $A^8$.  In the $(A^8,\widetilde{B})$ 
basis, the mass squared matrix is
\begin{equation}
(1+2x^2) M^2 \left(\begin{array}{cc} g_3^2/3  & -g_3\tilde{g}'/\sqrt{3} 
\\ -g_3\tilde{g}'/\sqrt{3} & \tilde{g}'^2 \end{array}\right).
\end{equation}
One immediately obtains the mass eigenstates
\begin{eqnarray}
B_L&=&c_\psi A^8 - s_\psi \widetilde{B} \\
B_H&=&s_\psi A^8 + c_\psi \widetilde{B} \,\,\,,
\end{eqnarray}
with
\begin{equation}
s_\psi=\frac{-g_3}{\sqrt{g_3^2+3\tilde{g}'^2}},\quad
\mbox{ and } \quad c_\psi=\frac{\sqrt{3}\tilde{g}'}{\sqrt{g_3^2+3\widetilde{g}'^2}}\,\,\,,
\end{equation}
and the masses
\begin{eqnarray}
M_{B_L}&=&0 \\
M_{B_H}&=& \left(1+2x^2\right)^{1/2}\left(\frac{g_3^2}{3}+\tilde{g}'^2\right)^{1/2}\,M.
\label{eq:bhmass}
\end{eqnarray}

Finally, the $\mathbf{2_{\pm 3/2}}$ state is formed from the remaining components of the SU(3) adjoint,
$A^a$ for $a=4,5,6,7$.  Its mass is given by
\begin{equation}
M_{3/2} =\frac{1}{\sqrt{2}}(1+x^2)^{1/2} g_3\, M,
\label{eq:xhmass}
\end{equation}
where the subscript indicates the hypercharge of the state.  For $g_3\ll \tilde{g},\tilde{g}'$, the
${\bf 2_{\pm 3/2}}$ gauge bosons will be significantly lighter than the other massive bosons, $W_H^a$ and $B_H$.

The remainder of the exotic particle spectrum originates from the Higgs fields $\Sigma$ and $\chi$.  Given 
the branching rule $\mathbf{3}=\mathbf{2_{1/2}} + \mathbf{1_{-1}}$ under the SU(2)$_0\times$U(1)$_0$ subgroup, 
one deduces the following SU(2)$_W\times$U(1)$_Y$ decompositions of the scalar fields:
\begin{equation}
\Sigma \rightarrow (\mathbf{2_{1/2}}+\mathbf{1_{-1}})\otimes \mathbf{2_{-1/2}} 
= \mathbf{3_0}+\mathbf{1_0}+\mathbf{2_{-3/2}}\,\,\,,
\end{equation}
\begin{equation}
\chi \rightarrow (\mathbf{2_{1/2}}
+\mathbf{1_{-1}})\otimes \mathbf{1_{1}} =\mathbf{2_{3/2}} + \mathbf{1_0}.
\end{equation}
Therefore, the symmetry-breaking sector consists of complex scalar fields in the following 
$SU(2)_W\times U(1)_Y$ representations
\begin{equation}
\mathbf{3_0}+\mathbf{1_0}+\mathbf{1_0}+\mathbf{2_{3/2}}+\mathbf{2_{-3/2}}.
\end{equation}
The spectrum of the scalar sector is model dependent.  Generically, one expects that all physical scalar
states should obtain masses of order the symmetry-breaking scale $M$.  Here we will verify this statement
by finding a local minimum of the scalar potential.

To construct a potential, we first list the possible gauge-invariant operators involving $\Sigma$ and $\chi$, 
up to quartic order in these fields.  We find
\begin{equation}
 t_1=m^2\Tr\Sigma^\dag\Sigma \,\,\,,
\end{equation}
\begin{equation}
 t_2= \Tr\Sigma^\dag\Sigma\Tr\Sigma^\dag\Sigma\,\,\,,
\end{equation}
\begin{equation}
 t_3=\Tr\Sigma^\dag\Sigma\Sigma^\dag\Sigma\,\,\,,
\end{equation}
\begin{equation}
 t_4=\Tr\Sigma \epsilon \Sigma^T \Sigma^* \epsilon \Sigma^\dag \,\,\,,
\end{equation}
\begin{equation}
 t_5=m^2 \chi^\dag \chi\,\,\,,
\end{equation}
\begin{equation}
 t_6=\chi^\dag \chi\chi^\dag \chi\,\,\,,
\end{equation}
\begin{equation}
t_7=\chi^\dag\Sigma\Sigma^\dag\chi\,\,\,,
\end{equation}
\begin{equation}
t_8=m \Sigma^i_{\alpha}\Sigma^j_{\beta}\epsilon^{\alpha\beta}\chi^{k}\epsilon_{ijk} + \mbox{ h.c.}\,\,\,,
\end{equation}
\begin{equation}
t_9=\chi^\dag \chi\Tr\Sigma^\dag\Sigma\,\,\,,
\end{equation}
where
\begin{equation}
\epsilon=\left(\begin{array}{cc} 0&1 \\ -1&0 \end{array}\right),
\end{equation}
and where $m$ is a mass of the same order as the desired symmetry-breaking scale.  The potential is
an arbitrary linear combination of these invariant terms, with coefficients $\alpha_i$,
\begin{equation}
V=\sum_{i=1}^9 \alpha_i\, t_i.
\label{eq:su3pot}
\end{equation}
To find a local minimum, we perform a constrained minimization.  Substituting  Eqs~(\ref{eq:mdef1}) 
and (\ref{eq:xdef1}) into the potential, we minimize the resulting function $V_0$, given by
\begin{equation}
 V_0=2\alpha_1 M^2+(4\alpha_2+2\alpha_3-2\alpha_4)M^4 + \alpha_5 M^2 x^2 
+\alpha_6 M^4 x^4 +4\alpha_8 M^3 x +2\alpha_9 M^4 x^2.
\end{equation}
For an example, setting $(\alpha_1,\ldots,\alpha_9) = (-1,1.1,1.2,1.4,-1.3,0.9,0.7,-0.8,0.5)$, we find that
the global minimum of the potential over the parameters $M$ and $x$ is at $(M,x)=(0.71994\,m, 1.32919)$. 
We confirm that this point is a minimum by studying the scalar mass squared matrix
\begin{equation}
 M_{ij}^2=\frac{\partial^2 V}{\partial \phi_i \partial \phi_j}  \,\,\,,
\end{equation}
where the $\phi_i$ denote the real scalar degrees of freedom in the fields $\Sigma$ and $\chi$, with
$1\le i \le 18$.  The squared masses are all positive, as shown in Table~\ref{table:su3sspec},
and have the correct multiplicity to occupy complete representations of the unbroken gauge group.
\begin{table}
\begin{center}
\caption{Spectrum of physical scalars in the SU(3) model, in units of $m$, for the example parameter 
choice described in the text.}
\label{table:su3sspec}
\vspace{1em}
\begin{tabular}{ccc}\hline\hline
state   & \quad & mass  \\  \hline
$\mathbf{3_{0}}$ &&  $2.06$ \\ 
$\mathbf{2_{3/2}}$ && $1.30$\\ 
$\mathbf{1_{0}}$ &&  $1.55$\\ 
$\mathbf{1_{0}}$ &&  $1.40$\\ 
$\mathbf{1_{0}}$ &&  $1.32$\\  
\hline\hline
\end{tabular}
\end{center}
\end{table}
For this choice of parameters $\alpha_i$, we also confirm that there are eight zero eigenvalues,
corresponding precisely to the $12-4$ broken generators in the spontaneous breaking 
SU(3)$\times$SU(2)$\times$U(1)$\rightarrow$SU(2)$_W\times$U(1)$_Y$.

\subsection{$G_U=$G$_2$}\label{sec:g2model}
We construct the model with $G_U=$G$_2$ in analogy to the SU(3) model of Section~\ref{sec:su3model}, 
though in this case we do not include vector-like matter.  The group G$_2$ contains SU(3) as a maximal 
subgroup; the fundamental representation of G$_2$ is seven dimensional
and decomposes under this SU(3) as
\begin{equation}
{\bf 7} = {\bf 3} + {\bf \bar{3}} + {\bf 1} \,\,\,.
\label{eq:seven}
\end{equation}
This result suggests the natural generalization of the Higgs sector of the
SU(3) theory:  Symmetry breaking is achieved via two fields, $\Sigma$ and $\chi$ with the
following quantum numbers and vevs:
\begin{equation}
\Sigma \sim ({\bf 7},{\bf {\bar{2}}_{-1/2}})  \,\,\,\,\, , \,\,\,\,\,
\langle \Sigma \rangle = \left(\begin{array}{cc} M & 0\\ 0 & M \\ 0 & 0 \\
0 & 0 \\0 & 0 \\0 & 0 \\0 & 0 \\
\end{array}\right)
\,\,\,\,\, , \,\,\,\,\, \chi \sim ({\bf 7},{\bf {1}_{1}})
\,\,\,\,\, , \,\,\,\,\,
\langle \chi \rangle = \left(\begin{array}{c} 0\\ 0 \\ x\,M \\
0 \\0 \\0 \\0 
\end{array}\right)\,\,\,.
\label{eq:g2chisigdef}
\end{equation}
With the vevs shown in Eq.~(\ref{eq:g2chisigdef}), G$_2$ is broken to SU(2)$_W\times$U(1)$_Y$.  
The adjoint representation of G$_2$ is 14-dimensional and decomposes as 
${\bf 8}+{\bf 3}+{\bf\bar{3}}$ under the SU(3) subgroup.  This immediately tells us that the 
electroweak quantum numbers of the gauge bosons are given by,
\begin{equation}
{\bf 14} = ({\bf 3_0}+{\bf 1_0}+{\bf 2_{3/2}}+{\bf 2_{-3/2}})+({\bf 2_{1/2}}+{\bf 1_{-1}})+
({\bf 2_{-1/2}}+{\bf 1_1}) \,\,\,.
\end{equation}
Again, pairs of representations with opposite hypercharges correspond to complex vector
fields.  With $M$ and $x$ defined as in Eq.~(\ref{eq:g2chisigdef}), the mass spectrum of 
$W_H$, $W_L$, $B_H$, $B_L$, and the exotic ${\bf 2_{\pm 3/2}}$ states, as well as the 
mixing angles $\phi$ and $\psi$, are precisely the same as in the SU(3) model of 
Section~\ref{sec:su3model}, with the identification 
\begin{equation}
g_3=g_2/\sqrt{2} \,\,\,,
\end{equation}
where $g_2$ is the G$_2$ gauge coupling.  We find that the masses of the new SU(2) doublet and 
singlet bosons are given by
\begin{equation}
M_{1/2} = \frac{1}{\sqrt{6}}(3+x^2)^{1/2}\, g_3 \, M \,\,\,,
\label{eq:m12mass}
\end{equation}
\begin{equation}
M_{1} = \frac{1}{\sqrt{3}} (1+x^2)^{1/2} \, g_3 \, M \,\,\,,
\label{eq:m1mass}
\end{equation}
where the subscripts again refer to the hypercharges of the states. Note that the ${\bf 1_{\pm 1}}$
states are always lighter than the ${\bf 2_{\pm 3/2}}$ bosons of the SU(3) model.  We postpone a 
discussion of the sector that facilitates the decays of these states until Section~\ref{sec:four}.

We now consider the scalar sector of the model.  It follows from the decompositions in
Eqs.~(\ref{eq:br3}) and (\ref{eq:seven}) that the $\Sigma$ and $\chi$ fields contain the 
following SU(2)$_W\times$U(1)$_Y$ reps:
\begin{equation}
\Sigma = ({\bf 3_0} + {\bf 1_0} + {\bf 2_{-3/2}}) + ({\bf 3_{-1}}+{\bf 1_{-1}}+{\bf 2_{1/2}})
+{\bf 2_{-1/2}} \,\,\,,
\end{equation}
\begin{equation}
\chi = ({\bf 2_{3/2}}+{\bf 1_0})+({\bf 2_{1/2}}+{\bf 1_2})+{\bf 1_{1}} \,\,\,.
\end{equation}
We will show that there are local minima of a $G_2$ invariant potential with the 
desired pattern of symmetry-breaking vevs.  Compared to the SU(3) model, however, the 
construction of invariant interactions is more interesting.  In constructing the potential for the 
SU(3) model, we used the fact that SU(2) has only real representations, so that
\begin{equation}
{\bf 2} \sim \epsilon \, {\bf 2}^*  \,\,\,.
\label{eq:ttb}
\end{equation}
where $\epsilon = i \,\sigma^2$. The group G$_2$ also has only real representations.   One finds 
that 
\begin{equation}
{\bf 7} \sim S \, {\bf 7}^* \,\,\,,
\label{eq:g2sim}
\end{equation}
where $S$ is the matrix
\begin{equation}
S = \left(\begin{array}{ccc} 
0 & \,\,\,\openone & 0 \\
\openone & \,\,\, 0 & 0 \\
0 & \,\,\, 0 & -1 
\end{array}\right) \,\,\,,
\end{equation}
and where $\openone$ is a three-by-three identity matrix.  Eq.~(\ref{eq:g2sim}) allows the construction 
of many more invariants compared to the potential in the SU(3) model.  Finally, the cubic invariant 
in the potential for the SU(3) model, which exists because it is possible to make a singlet out of 
three triplets, has a natural generalization in the G$_2$ model.  We note that the tensor product~\cite{atom}
\begin{equation}
{\bf 7}\times {\bf 7}= {\bf 1} + {\bf 7} + {\bf 14} + {\bf 27}  
\end{equation}
implies that ${\bf 7}^3$ also contains a singlet.  In SU(3), the appropriate Clebsch-Gordan
coefficients for making a singlet from three triplets is the three-index epsilon tensor
$\epsilon_{ijk}$;  in G$_2$, we find that the analogous object is a totally antisymmetric
tensor $C_{ijk}$, with
\begin{equation}
C_{714}=C_{725}=C_{736}=1 \,\,\,\,\, , \,\,\,\,\,
C_{123}=C_{456}=-\sqrt{2} \,\,.
\label{eq:cdef}
\end{equation}
All components that are not related to these by total antisymmetry are vanishing.  The nonvanishing
components of $C_{ijk}$ can be understood by considering the transformation properties of a cubic
invariant, for example
\begin{equation}
\psi_1^i \psi_2^j \psi_3^k C_{ijk} \,\,\,,
\label{eq:excube}
\end{equation}
where $\psi_i \sim {\bf 7}$, under the SU(3) subgroup.  Referring to the basis defined in Appendix~A, the 
components with one index equal to $7$ couple a ${\bf 3}$ in one ${\bf 7}$ to a ${\bf \bar{3}}$ in another 
and combine this with a singlet from the third.  The components with indices 123 (456) provide an SU(3)
epsilon tensor that couples three ${\bf 3}$'s (${\bf \bar{3}}$'s), with one originating from 
each of the ${\bf 7}$'s.  All the other elements of $C_{ijk}$ that are not related to these by
antisymmetry must be zero since they would not lead to a result that is invariant under the SU(3)
subgroup.  The relative normalization between the components that lead to different 
SU(3) invariants ({\em i.e.}, the factor of $-\sqrt{2}$ in Eq.~(\ref{eq:cdef})) is fixed by
the condition that Eq.~(\ref{eq:excube}) remain invariant under the action of the full group.  
We have checked this for infinitesmal transformations; our result for $C_{ijk}$ satisfies
\begin{equation}
C_{i'jk} (T^a)^{i'}_i+C_{ij'k} (T^a)^{j'}_j+C_{ijk'} (T^a)^{k'}_k=0
\end{equation}
for $a=1,\ldots,14$, where the $T^a$ are the G$_2$ generators given in the Appendix A.  

In analogy to Eq.~(\ref{eq:su3pot}), we may write the $\Sigma$-$\chi$ potential as
\begin{equation}
V = \sum_1^{12} \beta_i u_i  \,\,\, ,
\label{eq:g2potsum}
\end{equation}
where the $\beta_i$ are parameters and the $u_i$ are the following G$_2\times$SU(2)$\times$U(1)
invariants:
\begin{equation}
u_1 = m^2 \Tr \Sigma^\dagger \Sigma
\end{equation}
\begin{equation}
u_2 = \Tr \Sigma^\dagger \Sigma \,  \Tr \Sigma^\dagger \Sigma
\end{equation}
\begin{equation}
u_3=\Tr \Sigma^\dagger \Sigma \Sigma^\dagger \Sigma
\end{equation}
\begin{equation}
u_4=\Tr \Sigma^T S \Sigma \Sigma^\dagger S \Sigma^*
\end{equation}
\begin{equation}
u_5=\Tr \Sigma \epsilon \Sigma^T \Sigma^* \epsilon \Sigma^\dagger
\end{equation}
\begin{equation}
u_6 = \Tr \Sigma^T S \Sigma \epsilon \Tr \Sigma^\dagger S \Sigma^* \epsilon
\end{equation}
\begin{equation}
u_7=m^2 \chi^\dagger \chi
\end{equation}
\begin{equation}
u_8= \chi^\dagger \chi\chi^\dagger \chi
\end{equation}
\begin{equation}
u_9 = \chi^T S \chi \chi^\dagger S \chi^*
\end{equation}
\begin{equation}
u_{10}=m \Sigma^i_\alpha \Sigma^j_\beta \epsilon^{\alpha\beta} \chi^k \, C_{ijk} + \mbox{ h.c.}
\end{equation}
\begin{equation}
u_{11}=\chi^\dagger \Sigma \Sigma^\dagger \chi
\end{equation}
\begin{equation}
u_{12}=\chi^T S \Sigma \Sigma^\dagger S \chi^*
\end{equation}
Eq.~(\ref{eq:g2potsum}) is a function of $42$ real scalar degrees of freedom, and depends
on $12$ free parameters.  A complete study of this potential is beyond the scope of this paper. 
Nevertheless, we can again show that there are extrema of this potential with the desired 
symmetry-breaking vevs that correspond to local minima.   For this purpose, we first extremize
the potential with the constraint that the non-vanishing vevs are those displayed in
Eq.~(\ref{eq:g2chisigdef}); for these solutions we then determine whether the second-derivative
matrix of the potential is positive semi-definite, the same procedure that we employed in the SU(3) model. 
 As an example, for the choice of parameters
$(\beta_1 \ldots \beta_{12}) = (-1,0.3,1.1,1.0,1.0,1.0,-1.0,0.3,1.0,1.0,1.0,0.1)$, one obtains
$M=2.79\,m$, $x=1.25$ and the mass spectrum shown in Table~\ref{table:g2sspec}. This parameter choice
was random; generically, we don't find any fine-tuning is necessary to find solutions.
\begin{table}
\begin{center}
\caption{Spectrum of physical scalars, in units of $m$, for the example parameter choice described
in the text.}
\label{table:g2sspec}
\vspace{1em}
\begin{tabular}{ccccccc}\hline\hline
state  &\quad & mass  &\quad\quad &state &\quad & mass \\ \hline
${\bf 3_{-1}}$  & & $4.52$ && ${\bf 1_2}$ & & $5.24$ \\
${\bf 3_0}$     & & $6.52$ && ${\bf 1_1}$ & & $2.84$  \\
${\bf 2_{3/2}}$ & & $4.25$ && ${\bf 1_0}$ & & $4.03$ \\
${\bf 2_{1/2}}$ & & $3.03$  && ${\bf 1_0}$ & & $3.61$ \\
${\bf 2_{1/2}}$ & & $2.19$  && ${\bf 1_0}$ & & $2.25$ \\
\hline\hline
\end{tabular}

\end{center}
\end{table}
Since we have established that there is no difficulty in finding appropriate symmetry-breaking 
vacua in both the SU(3) and G$_2$ models, we take $M$ and $x$ as free parameters in
the phenomenological analysis that follows.  

\section{Electroweak Constraints}\label{sec:three}

The most important electroweak constraints on models with the symmetry-breaking 
$G_U\times$SU(2)$\times$U(1)$\rightarrow$SU(2)$_W\times$U(1)$_Y$ comes from the tree-level
mixing between the SU(2)$_0\times$U(1)$_0$ and the SU(2)$\times$U(1) gauge bosons, where
SU(2)$_0\times$U(1)$_0 \subset G_U$.  The gauge bosons corresponding to the broken
generators of the original gauge group do not couple directly to Standard Model matter 
fields, aside from higher-dimension operators that are generically suppressed.  For this reason, 
the analysis of electroweak constraints on the SU(3) and G$_2$ models of interest to us here proceeds
 in analogy to that of the minimal weak SU(3) model discussed in Ref.~\cite{cekt}.  The present analysis 
differs from that of the minimal SU(3) model in three ways: ({\em i.}) The low-energy effective Lagrangian
that one obtains after integrating out the heavy $W_H$ and $B_H$ fields depends on their masses,
which differ in our model due to the $\chi$ field vev.  Our results therefore depend on the 
additional parameter $x$ defined in Eqs.~(\ref{eq:xdef1}) and (\ref{eq:g2chisigdef}) ({\em ii.}) The presence
of the $\chi$ field, as well as vector-like matter in the SU(3) model, alters the beta functions that determine 
the matching between the low-energy gauge couplings, $g$ and $g'$, and the couplings of the high-energy 
theory, $g_U$, $\tilde{g}$ and $\tilde{g}'$, via Eq.~(\ref{eq:match}). ({\em iii.}) We use updated data for
the electroweak observables that was not available at the time of Ref.~\cite{cekt}.

\begin{figure}
\includegraphics[scale=.6]{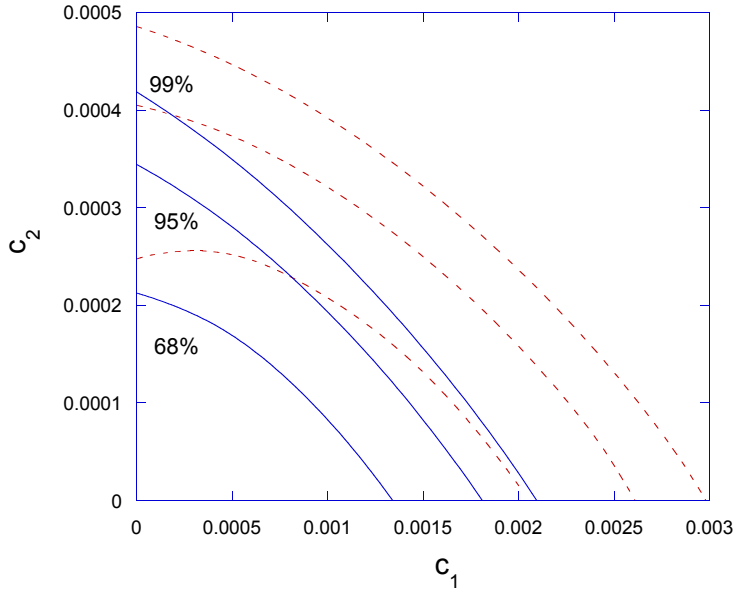}
\caption{Confidence level contours in the $c_1$-$c_2$ plane for the electroweak fit
described in the text.  The dotted lines show the corresponding results for the input
data used in Ref.~\cite{cekt}.} \label{fig:chicomp}
\end{figure}

Since our approach follows that of Ref.~\cite{cekt}, we will confine our discussion to the
way in which the analysis is modified.  Letting $\Phi$ represent the Higgs field of the minimal
Standard Model, corrections to the ordinary $W$ and $Z$-boson masses follow from study of the kinetic
term $(D_\mu\Phi)^\dagger(D^\mu \Phi)$, which includes couplings to both the $W_H$, $W_L$, $B_H$ and
$B_L$ fields.  The heavy fields can be eliminated by applying their equations of motion so that
all the quadratic terms can be expressed in terms of the light fields only.  Following this approach, we
find
\begin{equation}
m_W^2 = \frac{g^2 v^2}{4} (1 - \frac{v^2 c_\phi^4}{4 M^2})  \,\,\,,
\label{eq:wm}
\end{equation}
\begin{equation}
m_Z^2 = \frac{v^2}{4} (g^2+{g'}^2) \left[1-\frac{v^2}{4 M^2}(c_\phi^4+\frac{1}{1+2\,x^2} 
c_\psi^4)\right] \,\,\,,
\label{eq:zm}
\end{equation}
where  $\langle \Phi \rangle = v/\sqrt{2}$ with $v=246$~GeV.  Similarly, exchanges of the $W_H$ and $B_H$ 
fields alter the couplings of the Standard Model gauge fields to ordinary matter.  Since Standard Model 
fields couple only to the extra SU(2)$\times$U(1) factors, the gauge-matter couplings are summarized by
\begin{equation}
{\cal L}_J = \tilde{g}\, \tilde{W}^a_\mu J^{a\mu} + \tilde{g}' \tilde{B}_\mu J_Y^{\mu} \,\,\,,
\end{equation}
where $J^a$ and $J_Y$ are the conventional SU(2)$_W$ and U(1)$_Y$ currents of the Standard Model.  
The fields $\tilde{W}$ and $\tilde{B}$ may be expressed in terms of $W_H$, $W_L$, $B_H$ and
$B_L$, and the heavy fields can again be eliminated by use of their equations of motion.  We find
\begin{eqnarray}
{\cal L} &=& g \, \left(1-\frac{v^2}{4 M^2}c_\phi^4 \right) W_L^a \cdot J^a + g'\left(1-\frac{v^2}{4 M^2}
\frac{c_\psi^4}{1+2\, x^2}\right) B_L \cdot J_Y 
+ g'\left(\frac{v^2c_\phi^4}{4 M^2}\right) B_L \cdot J^3 \nonumber \\
&+& g \,\left(\frac{v^2}{4 M^2}\frac{c_\psi^4}{1+2\, x^2}\right) 
W_L^3 \cdot J_Y  -\left(\frac{c_\phi^4}{2 M^2}\right) J^a \cdot J^a
-\left(\frac{1}{2 M^2} \frac{c_\psi^4}{1+2\, x^2}\right) J_Y \cdot J_Y  \,\,\,.
\label{eq:currents}
\end{eqnarray}
The leading electroweak corrections in our models follow from Eqs.~(\ref{eq:wm}), (\ref{eq:zm}) and 
(\ref{eq:currents}).  These results agree with the corresponding expressions in Ref.~\cite{cekt} when $x=0$,
and indicate that the theoretical prediction for any electroweak observable can be obtained from the $x=0$ 
result via the substitution
\begin{equation}
c_\psi^4 \longrightarrow \frac{c_\psi^4}{1+2 \, x^2} \,\,\,.
\end{equation}
Theoretical predictions for 22 electroweak observables  are given in an Appendix of Ref.~\cite{cekt}, 
and are expressed as a function of two parameters, $c_1$ and $c_2$. These predictions remain valid in our 
models if we identify
\begin{equation}
c_1 = \frac{c_\phi^4 v^2}{4 M^2} \,\,\,\,\,\mbox{ and } \,\,\,\,\, 
c_2= \frac{1}{1+2\,x^2}\frac{c_\psi^4 v^2}{4 M^2} \,\,\,\,.
\end{equation}
Following the approach of Ref.~\cite{cekt}, we construct a chi-squared function for the
shifts in electroweak precision observables from their Standard Model values as a function 
of the parameters $c_1$ and $c_2$.  The Standard Model predictions and experimental data
are taken from a fit by Langacker and Erler that appears in the 2006 Review of Particle 
Properties~\cite{rpp}.  For convenience, we quote these values in Table~\ref{table:ewdata}.
\begin{table}
\begin{center}
\caption{Input parameters for the electroweak fit described in the text. The SM column shows
central values from a Standard Model fit by Langacker and Erler, appearing in the 2006 
Review of Particle Physics~\cite{rpp}, in which $m_Z=91.1874\pm 0.0021$~GeV,
$m_H=89^{+38}_{-28}$~GeV, $m_t=172.7\pm2.8$~GeV and $\alpha_s(m_Z)=0.1216\pm0.0017$.}
\label{table:ewdata}
\vspace{1em}
\begin{tabular}{ccccccc}\hline\hline
 Quantity  & Experiment  &  SM &\quad&  Quantity   & Experiment  & SM\\ \hline
$\Gamma_Z$  & $2.4952\pm0.0023$ & 2.4968 && $A_e(P_\tau)$    & $0.1498\pm0.0049$ & 0.1471\\
$R_e$       & $20.8040\pm0.0500$ & 20.7560&& $A_{FB}^b$       & $0.0992\pm0.0016$ & 0.1031\\
$R_\mu$     & $20.7850\pm0.0330$ & 20.7560&& $A_{FB}^c$       & $0.0707\pm0.0035$ & 0.0737\\
$R_\tau$    & $20.7640\pm0.0450$ & 20.8010&& $A_{LR}$         & $0.15138\pm0.00216$ & 0.1471\\
$\sigma_h$  & $41.5410\pm0.0370$ & 41.4670&& $M_W$ & $80.403\pm0.029$ & 80.3760\\
$R_b$       & $0.21629\pm0.00066$ & 0.21578&& $M_W/M_Z$ & $0.88173\pm0.00032$ & 0.8814\\
$R_c$       & $0.1721\pm0.0030$ & 0.17230&& $g_L^2(\nu N\rightarrow \nu X)$ & $0.30005\pm0.00137$ & 0.30378\\
$A_{FB}^e$  & $0.0145\pm0.0025$ & 0.01622&& $g_R^2(\nu N\rightarrow \nu X)$ & $0.03076\pm0.00110$ & 0.03006\\  
$A_{FB}^\mu$ & $0.0169\pm0.0013$ & 0.01622&& $g_{eA}(\nu e\rightarrow \nu e)$ & $-0.5070\pm0.014$ & -0.5064\\
$A_{FB}^\tau$    & $0.0188\pm0.0017$ & 0.01622&& $g_{eV}(\nu e\rightarrow \nu e)$ & $-0.040\pm0.015$ & -0.0396\\
$A_\tau(P_\tau)$ & $0.1439\pm0.0043$ & 0.1471&& $Q_W(Cs)$ & $-72.62\pm0.46$ & -73.17\\
\hline\hline
\end{tabular}
\end{center}
\end{table}
We compute confidence contours by taking the new theory to be the null hypothesis. Then 
$\chi^2-\chi^2_{min}$ is also $\chi^2$ distributed, with two degrees of freedom (the number 
of parameters, $c_1$ and $c_2$).  The main difference in the electroweak data that we use in for 
our fit compared to Ref.~\cite{cekt} is that more recent LEP II results have shifted the central 
value of the $W$ mass downward.  Since the nonstandard contribution to $M_W$ in our models 
is positive, the parameter space is now more tightly constrained.  We illustrate this in 
Fig.~\ref{fig:chicomp}, which displays the 68\%, 95\% and 99\% confidence contours from our 
global fit compared to those in Ref.~\cite{cekt}.  The shift in these contours does 
not lead to a dramatic change in the allowed parameter space of the minimal SU(3) model.  The
shape of the exclusion region in our models depends more noticeably on the value of the parameter $x$, as
we describe below. 

The parameter space of the model may be described in terms of the couplings $\widetilde{g}'$,
and $\widetilde{g}$.  As in Ref.~\cite{cekt}, we define the unification scale $M_U$ as the mass of the 
heaviest gauge boson, the threshold at which the matching conditions Eq.~(\ref{eq:match}) should be 
applied. The Standard Model gauge couplings $g(M_U)$ and $g'(M_U)$, are determined via the one-loop
renormalization group equations
\begin{equation}
\alpha^{-1}(M_U) = \alpha^{-1}(M_Z) + \frac{b_{SM}}{2\pi}\ln\frac{M_U}{M_Z} + 
\sum_i \frac{b_i}{2\pi} \ln\frac{M_U}{M_i} \,\,\,,
\label{eq:rge1}
\end{equation}
\begin{equation}
{\alpha'}^{-1}(M_U) = {\alpha'}^{-1}(M_Z) + \frac{b'_{SM}}{2\pi}\ln\frac{M_U}{M_Z} + 
\sum_i \frac{b'_i}{2\pi} \ln\frac{M_U}{M_i}\,\,\,,
\label{eq:rge2}
\end{equation}
where $M_i$ is the mass of the $i^{th}$ heavy particle threshold, and $b_i$ the contribution 
to the beta function.  For the heavy gauge bosons, the $M_i$ are proportional to $M_U$, since
the unification scale is identified as the larger of Eq.~(\ref{eq:whmass}) or (\ref{eq:bhmass}); the
other heavy boson states are always lighter than this result.  The physical scalar components of 
the $\Sigma$ and $\chi$ Higgs fields are taken to have the same mass as the ${\bf 2_{3/2}}$ gauge bosons, 
the same approximation used in Ref.~\cite{cekt}.  When vector-like matter is included, the mass scale
is separately specified. The values for the beta functions are given in 
Table~\ref{table:betas}. 
\begin{table}[hbt]
\begin{center}
\begin{tabular}{ccccc}\hline\hline
         & \multicolumn{2}{c}{SU(3)} & \multicolumn{2}{c}{G$_2$} \\
states   & $b_i$ & $b'_i$ & $b_i$ & $b'_i$  \\ \hline
Standard Model &  $19/6$ & $-41/6$ & $19/6$ & $-41/6$ \\
${\bf 1_{\pm 1}}$   vector & - & - & $0$ & $7$ \\
${\bf 2_{\pm 1/2}}$ vector & - & - & $7/2$ & $7/2$ \\
${\bf 2_{\pm 3/2}}$ vector & $7/2$ & $63/2$  & $7/2$ & $63/2$\\
physical scalars  & $-1/2$ & $-3/2$ & $-3/2$ & $-9/2$ \\
vector-like & $-2 n_F/3$ & $-2 n_F$ & - & - \\ \hline\hline
\end{tabular}
\caption{Beta functions $b_i$ in Eqs.~(\ref{eq:rge1}) and (\ref{eq:rge2}).  Vector boson beta functions 
include the contribution from the longitudinal (eaten scalar) component.  Physical scalars are assumed to 
have the same mass as the ${\bf 2_{3/2}}$ vector bosons.  The number of $\mathbf{3}+\mathbf{\bar{3}}$ pairs
is given by $n_F$.}
\label{table:betas}
\end{center}
\end{table}
If one specifies $\widetilde{g}'$ and $\widetilde{g}$, then Eqs.~(\ref{eq:match}), (\ref{eq:rge1}) 
and (\ref{eq:rge2}) completely determine $M_U$ and the coupling $g_3(M_U)$.  The quantities 
$c_\phi$ and $c_\psi$ follow immediately, while the parameter $M$ is known through the identification 
of $M_U$ with the heaviest gauge boson mass.  All the quantities needed to compute the values of $c_1$ and 
$c_2$ are thereby obtained.  We implement this procedure numerically to associate each point in the 
$\widetilde{g}'$-$\widetilde{g}$ plane with a point in $c_1$-$c_2$ space; in this way, we determine whether 
a given point in the model's parameter space is excluded, to any desired confidence level.  We show the
95\% confidence level exclusion regions in the results that follow.

\begin{figure}
\includegraphics[scale=.6]{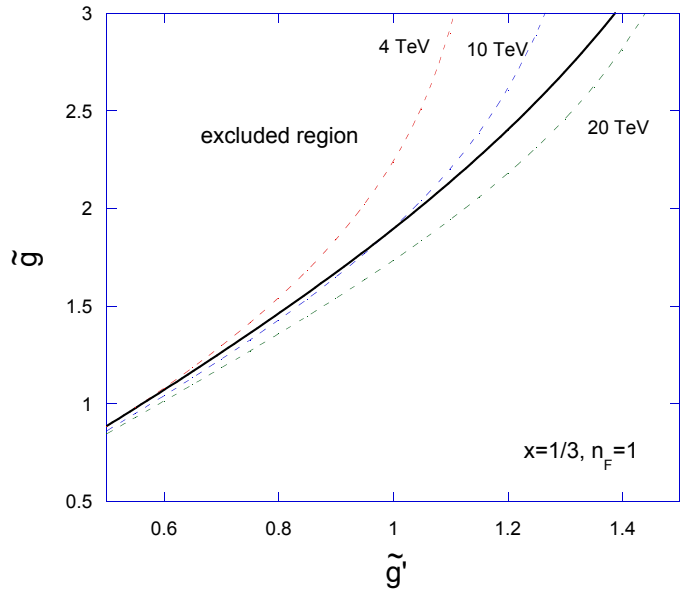}
\caption{Contours of the unification scale $M_U$ in the SU(3) model for
$x=1/3$, $n_F=1$ and $M_F=1$~TeV. The 95\% C.L. exclusion region is also shown.}\label{fig:x13n1}
\end{figure}
\begin{figure}
\includegraphics[scale=.6]{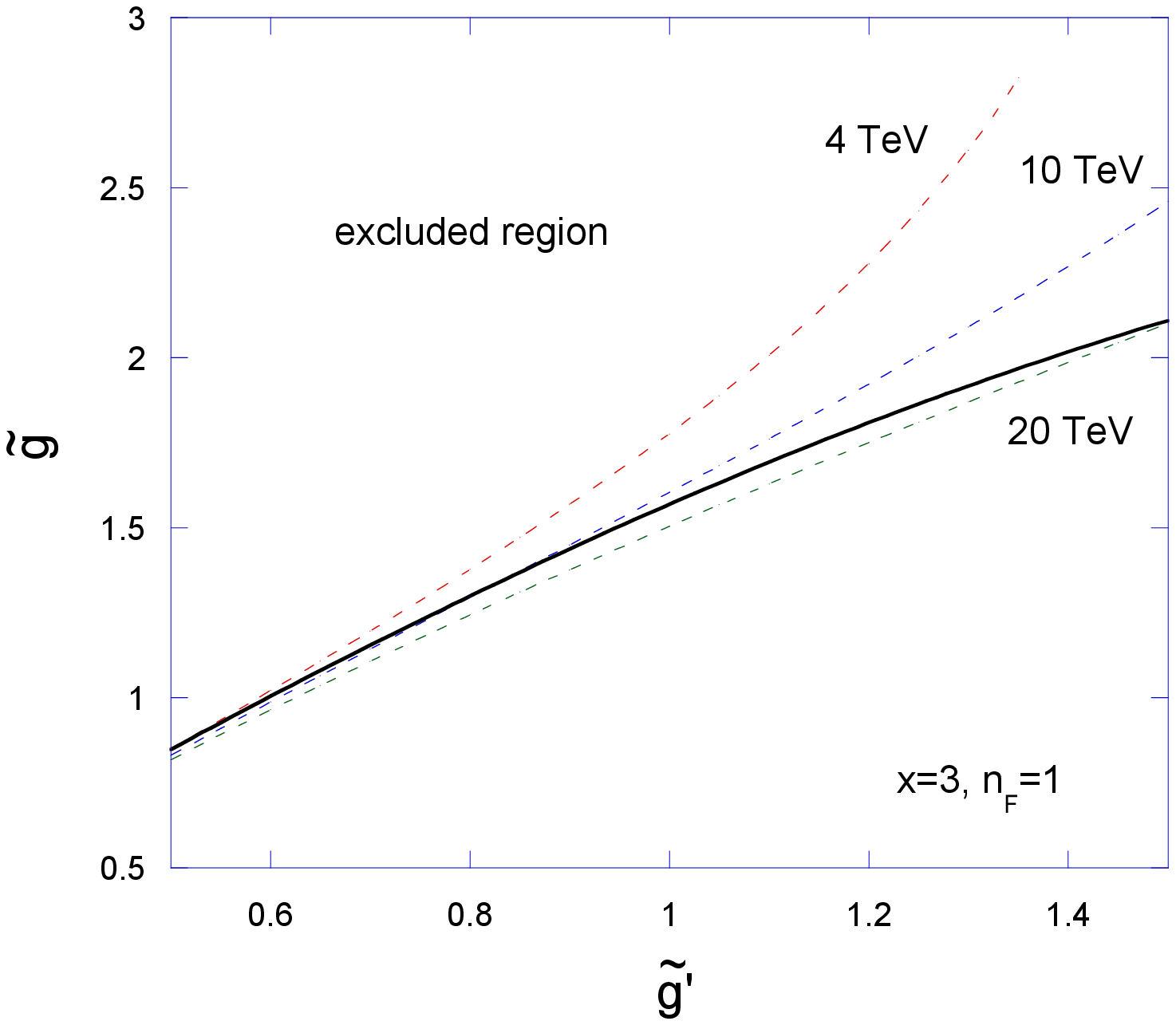}
\caption{Contours of the unification scale $M_U$ in the SU(3) model for
$x=3$, $n_F=1$ and $M_F=1$~TeV. The 95\% C.L. exclusion region is also shown.}\label{fig:x3n1}
\end{figure}
\begin{figure}
\includegraphics[scale=.6]{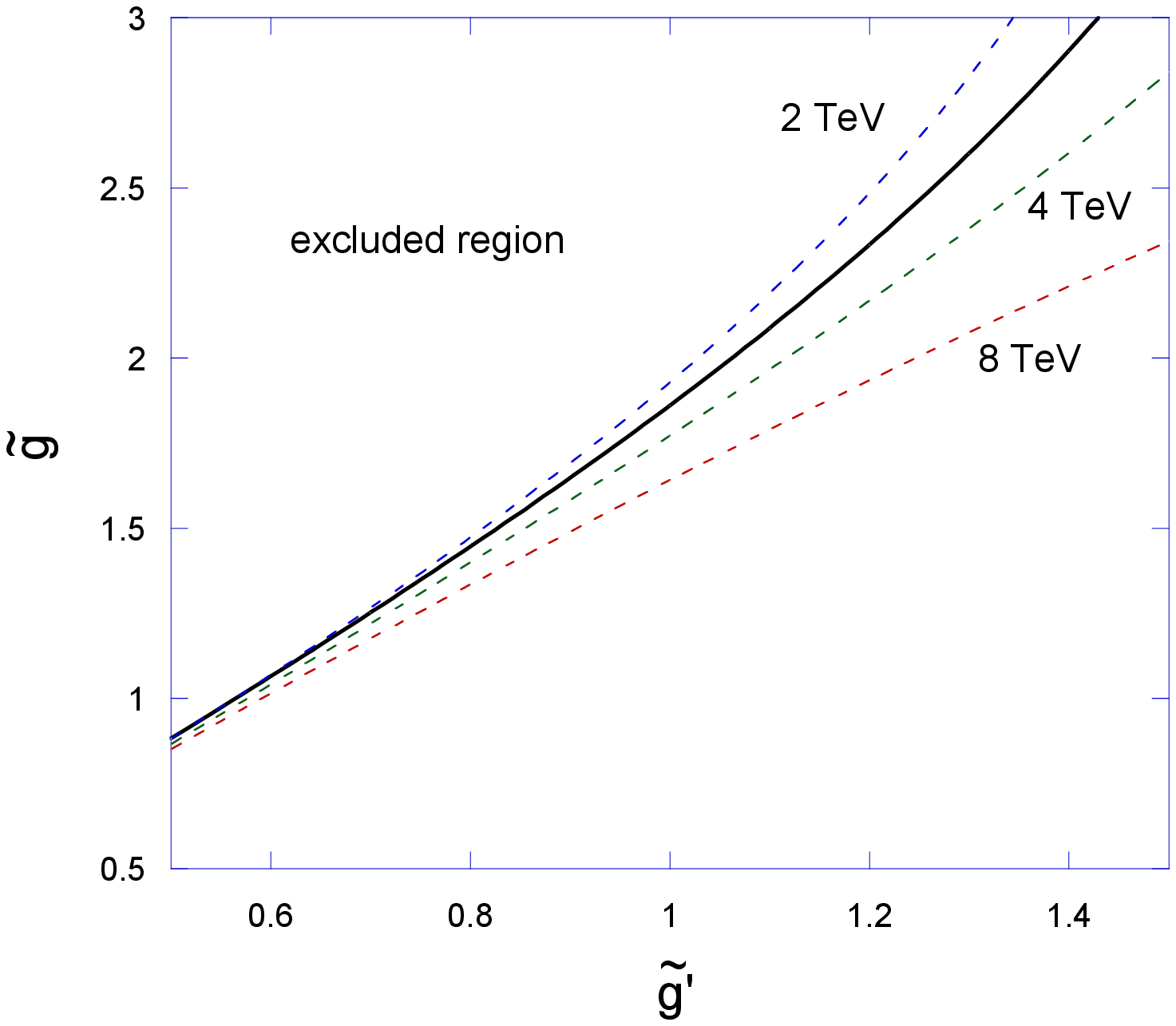}
\caption{Contours for $M_{3/2}$, the mass of lightest exotic gauge boson in the
SU(3) weak model, assuming $x=1$ and $n_F=0$. The 95\% C.L. exclusion region is also shown.}\label{fig:mxfig}
\end{figure}
To complete the analysis, we must specify values of $x$ and, in the SU(3) model, the number of 
${\bf 3}+{\bf \overline{3}}$  pairs $n_F$.  Figures~\ref{fig:x13n1} and \ref{fig:x3n1} show our
results for the matching scale $M_U$ assuming $n_F=1$ and $M_F=1$~TeV, allowing $x$ to vary between 
$1/3$ and $3$. Larger $x$ tends to exclude smaller values of $\tilde{g}$; however, the constant
$M_U$ contours and the boundary of the excluded region move in tandem, so that the effect on the smallest
allowed value of $M_U$ is relatively mild.  It is worth noting that there is an optimal choice
$x\approx1.2$ for which the $M_U=10$~TeV contour is within the allowed region for $\widetilde{g}'<1.2$ and 
$\widetilde{g}<2$, an improvement over the minimal SU(3) model. However, at large values, $x\ge 3$, the 
exclusion region grows, engulfing the entire $M_U = 10$~TeV contour. Varying the number of heavy 
fermion pairs between $0$ and $3$ has a negligible effect on the position of these contours or the 
excluded region, so we do not provide separate plots.   Figure~\ref{fig:mxfig} shows mass contours for
the ${\bf 2_{3/2}}$ gauge boson in the SU(3) model, with $n_F=0$.  Generally, we note that $M_{3/2} = 2$~TeV 
is entirely excluded and $M_{3/2} = 4$~TeV entirely outside the exclusion region.  For an optimal value 
of $x\approx 0.9$, the $M_{3/2} = 3$~TeV curve is completely outside the exclusion region.  Again, for 
$x\ge 3$, the exclusion region becomes large, excluding $M_{3/2}=4$~TeV as well\footnote{In all our 
figures, $\widetilde{\alpha}<1$ and $\widetilde{\alpha}'<1$ so that our analysis does not extend into the 
parameter space in which nonperturbative effects ({\em e.g.,} the formation of fermion condensates) may become 
important.  The ranges in $\widetilde{g}$ and $\widetilde{g}'$ shown were chosen to coincide with those of 
Ref.~\cite{cekt}, for ease of comparison.}.

\begin{figure}
\includegraphics[scale=.6]{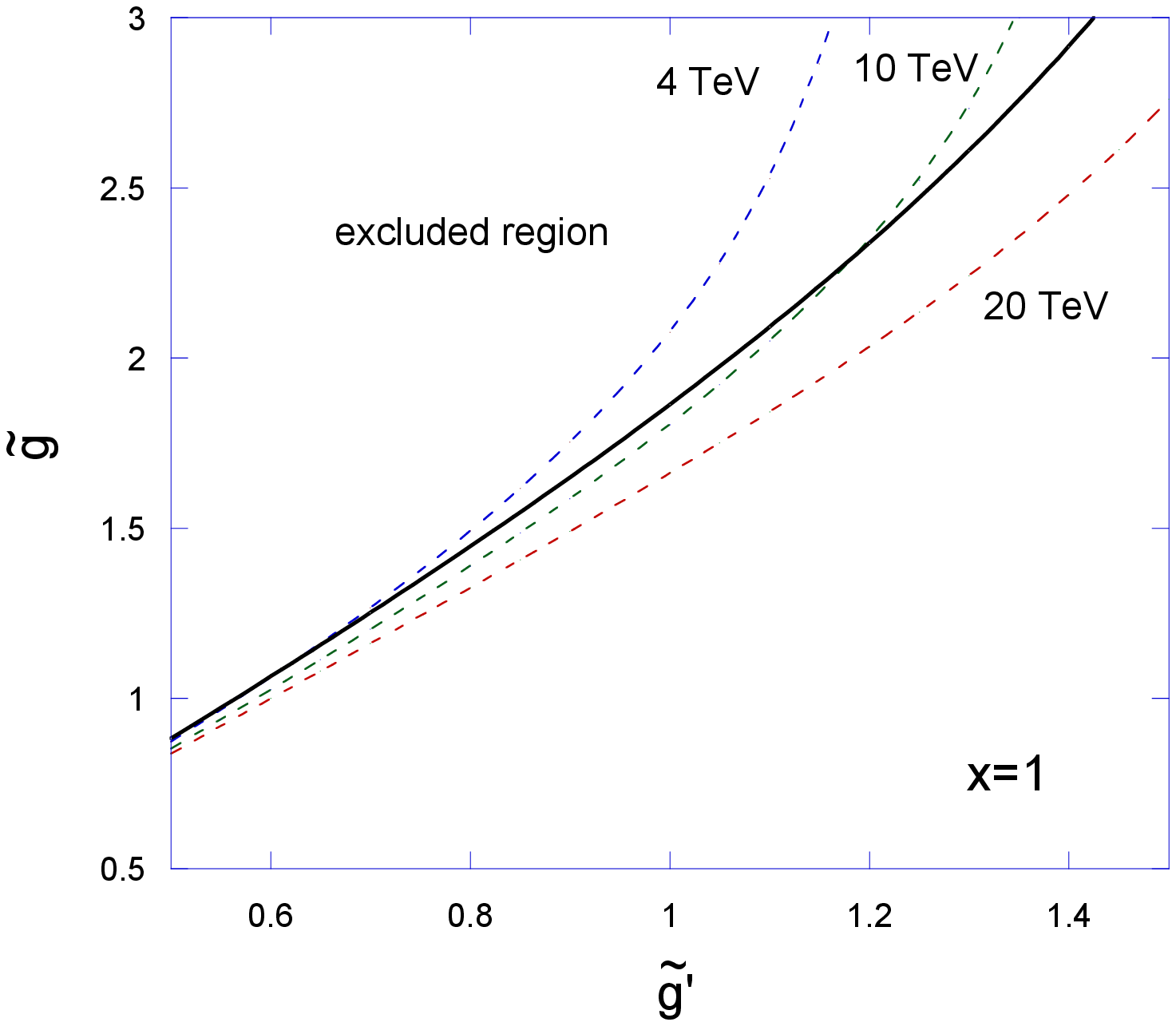}
\caption{Contours of the unification scale $M_U$ in the G$_2$ model for
$x=1$. The 95\% C.L. exclusion region is also shown.}\label{fig:g2mu}
\end{figure}
\begin{figure}
\includegraphics[scale=.6]{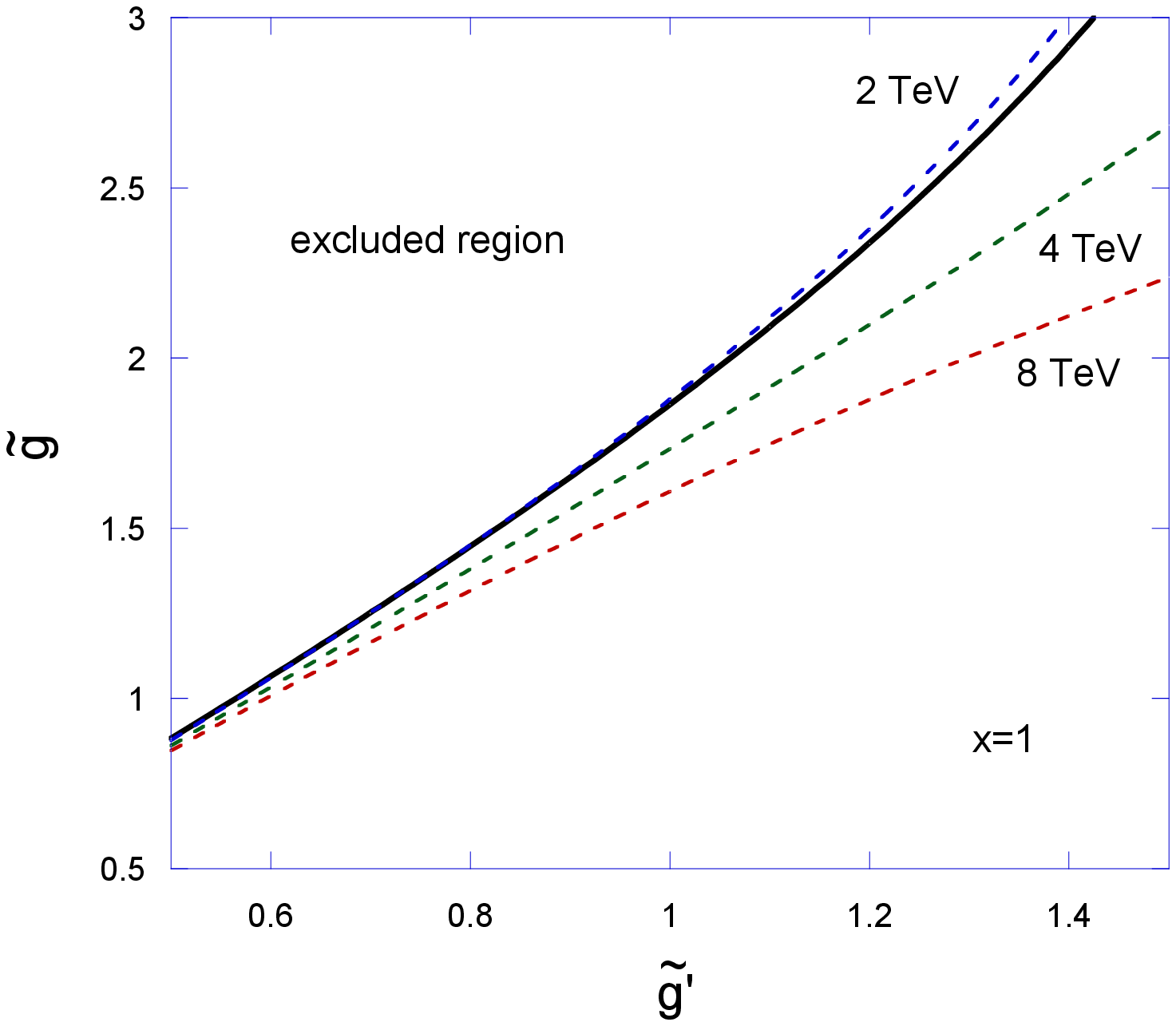}
\caption{Contours for $m_{1}$, the mass of lightest exotic gauge boson in the
G$_2$ model, assuming $x=1$.  For this choice, $m_{1}=m_{1/2}$. The 95\% C.L. exclusion region is 
also shown. }\label{fig:m1cont}
\end{figure}
In the case of the G$_2$ model, we work with $n_F=0$ since we will arrange for boson decays via higher-dimension
operators that do not necessarily arise via the exchange of vector-like matter.  The renormalization
group analysis differs since there are now additional mass thresholds, as indicated in Table~\ref{table:betas}.  
Note that the lightest gauge boson state in the G$_2$ model depends on the choice of parameters.  With $x \le 1$, 
the $\mathbf{1_{\pm 1}}$ field is the  lightest; its mass is always smaller by a factor of $\sqrt{2/3}$ 
relative to the  ${\bf 2_{\pm3/2}}$ state in the SU(3) model.   With $x > 1$, the $\mathbf{2_{\pm 1/2}}$ is 
the lighter, while for $x=1$ the $\mathbf{1_{\pm 1}}$ and $\mathbf{2_{\pm 1/2}}$ are degenerate.  
Figures~\ref{fig:g2mu} and \ref{fig:m1cont} show the unification and lightest exotic gauge boson contours,
respectively, for the choice $x=1$.  As one might anticipate, the results are qualitatively similar to
the SU(3) model, aside from the lighter exotic gauge boson states.  The fact that the $m_1=m_{1/2}=2$~TeV 
contour in Fig.~\ref{fig:m1cont} is almost contiguous with the exclusion line suggests that these exotic 
gauge bosons could be pair produced at the LHC. Ranges of minimum allowed gauge boson masses 
for approximately optimal values of $x$ are shown in Table~\ref{table:opt}.
\begin{table}
\begin{center}
\caption{Ranges of minimum allowed gauge boson masses in the G$_2$ model as $\tilde{g}'$ varies 
from $0.5$ to $1.5$, for a choice of $x$ near its optimal value. The result in the SU(3) model for the
$\mathbf{2_{\pm 3/2}}$ state is the same as in the G$_2$ model, to the accuracy shown.}
\label{table:opt}
\begin{tabular}{ccccc}\hline\hline
State &\quad &  $x$ &\quad& Mass range (TeV) \\ \hline
$\mathbf{2_{\pm 3/2}}$ && 0.9 && 2.27\,-\,2.75 \\
$\mathbf{1_{\pm 1}}$ && 0.9 && 1.85\,-\,2.22 \\ 
$\mathbf{2_{\pm 1/2}}$ && 1.5 && 1.76\,-\,2.15 \\
\hline\hline
\end{tabular}

\end{center}
\end{table}
The differing gauge boson spectrum of the G$_2$ model may therefore make it an easier candidate for direct 
detection at future collider experiments\footnote{Note that we do not exclude any additional region of the parameter space in our models
from the consideration of neutral meson mixing.  For example, the $W_H$ contribution to the 
$K_L$-$K_S$ mass splitting is suppressed relative to the Standard Model result, by a factor 
that is well approximated by $2(\tilde{g} \cos\phi/g)^2(M_W/M_{W_H})^2$, and does not exceed Standard Model hadronic 
uncertainties anywhere in the region allowed by the precision electroweak constraints.}.

\section{Decays}\label{sec:four}

In the minimal SU(3) model, the ${\bf 2_{\pm 3/2}}$ gauge bosons have no direct coupling to
Standard Model particles.  To avoid unwanted, charged stable states, the model presented in
Section~\ref{sec:su3model} included vector-like ${\bf 3}+{\bf \bar{3}}$ pairs. The
${\bf 3}$ has the same SU(2)$_W\times$U(1)$_Y$ decomposition as a lepton generation in the Standard
Model.  Mass mixing between the vector-like and Standard Model leptons will occur if there are Higgs fields
with the quantum numbers $({\bf 3},{\bf \bar{2}_{1/2}})+({\bf 3}, {\bf 1_{-1}})$ and with vevs in the components that 
transform as a ${\bf 1_0}$ under the SU(2)$_W\times$U(1)$_Y$ subgroup.  This corresponds precisely to our 
choice in Eqs.~(\ref{eq:mdef1}) and (\ref{eq:xdef1}).  

In this section, we will consider the case where the scale of the vector-like matter is higher than 
the unification scale, $M_F > M_U$.  In this limit, the vector-like matter may be 
integrated out of the theory, which simplifies the discussion of the heavy gauge boson decays.
We construct a low-energy effective Lagrangian following the same procedure described in 
Section~\ref{sec:three}. For definiteness, let us assume $n_F=1$. Starting with the Lagrangian 
in Eq.~(\ref{eq:hlmix}), one may read off the equation of motion for $\psi$ (or $\overline{\psi}$).  
To lowest order in $\not\!\!D/M_F$, the equation of motion for $\psi$ is solved by
\begin{equation}
\psi= -\frac{1}{M_F} \left(1+\frac{i \not\!\!D}{M_F}\right) \left[ \Sigma \lambda^\ell \ell_R
+\chi \lambda^e e_R\right] \,\,\,.
\end{equation}
Substituting this into Eq.~(\ref{eq:hlmix}) and discarding higher-order terms leads to the effective
Lagrangian
\begin{equation}
{\cal L}_{eff} = \frac{1}{M_F^2}(\overline{e}_R {\lambda^e}^\dagger \chi^\dagger) i \not\!\!D (\Sigma \lambda^\ell
\ell_R) +\frac{1}{M_F^2} (\overline{e}_R {\lambda^e}^\dagger \chi^\dagger) i \not\!\!D (\chi \lambda^e e_R) 
+\mbox{ h.c.} \,\,\, .
\label{eq:edl}
\end{equation}
If we had never mentioned vector-like matter, we could instead have postulated that new physics at a 
cut off scale $M_F$ generates higher-dimension operators, including those that lead to exotic gauge boson 
decays.  Symmetry and power-counting would have led us to Eq.~(\ref{eq:edl}) directly, with perhaps a 
different label for the unknown couplings in flavor space.  It is worth noting that both terms in 
Eq.~(\ref{eq:edl}) involve the field $\chi$ which is not present in the minimal SU(3) model; this leads to 
gauge boson decay via operators of much lower order than those mentioned in Ref.~\cite{cekt}.

Setting either some or all of the Higgs fields in Eq.~(\ref{eq:edl}) to their vevs leads to operators that
contribute to the decays of the new scalar or vector states in the model.  We focus the present discussion
on the exotic gauge bosons.  While one could imagine similar theories with different symmetry breaking sectors,
the gauge boson content of both our SU(3) and G$_2$ models are, by definition, model independent.   The 
interesting portion of the gauge boson matrix in the SU(3) model can be represented as
\begin{equation}
A^a T^a \supset \left(\begin{array}{cc} {\bf 0} & X \\ X^\dagger  &0
\end{array}\right) \,\,\, ,
\end{equation}
where $X \sim {\bf 2_{3/2}}$.  Substituting into Eq.~(\ref{eq:edl}), one may extract the $X$-fermion-fermion
vertex:
\begin{equation}
{\cal L}_X = g_3 x \left(\frac{M}{M_F}\right)^2 \overline{e}_R [{\lambda^e}^\dagger 
\not\!\!X^\dagger \lambda^\ell] \ell_R + \mbox{ h.c.}  \,\,\,.
\end{equation}
Notice that the coupling vanishes when $M_F \rightarrow \infty$ or $x\rightarrow 0$. 

In the G$_2$ model, the ${\bf 2_{\pm 3/2}}$ gauge bosons also decay as a consequence of the
operators in Eq.~(\ref{eq:edl}); the implicit SU(3) indices running from $1$ to $3$ in this
expression should simply be promoted to G$_2$ indices running from $1$ to $7$.  In the case
of the G$_2$ model, we will not assume the presence of vector-like matter, and pursue the simpler
approach of studying possible operators that may arise at a cut off $M_F$.   The operators in
Eq.~(\ref{eq:edl}) do not allow the new ${\bf 2_{\pm 1/2}}$ and ${\bf 1_{\pm 1}}$ G$_2$ gauge 
bosons to decay to Standard Model and/or neutral particles.  We find that the simplest way to remedy 
this is to introduce a new singlet fermion $\nu_R$, with the possible effective interactions
\begin{equation}
{\cal L}_{eff} = \frac{1}{M_F^2} \overline{\nu^c_R} \left( \chi^T S \right) \not\!\!D \left(\Sigma \ell_L
\right) + \frac{1}{M_F^2} \overline{\nu^c_R} \Tr \left[\Sigma^T S \not\!\!D \Sigma \epsilon\right] e_R^c
+ \mbox{ h.c.} \,\,\,,
\label{eq:rhn}
\end{equation}
where $S$ and $\epsilon$ are the matrices that were used in constructing gauge invariant operators
in Section~\ref{sec:g2model}.  In this expression, we let $\ell_L=(\nu_L,e^-_L)$ represent the Standard
Model lepton doublet with hypercharge $-1/2$. Working in the basis of G$_2$ generators given in the Appendix, 
we identify
\begin{equation}
A_{\bf 2_{1/2}} = \left[\begin{array}{c} (A^{13}+i A^{14})/\sqrt{2} \\ (A^{11}+i A^{12})/\sqrt{2} 
\end{array}\right] \,\,\,\,\,\,\,\,\,\, A_{\bf 1_{-1}}=(A^9+i A^{10})/\sqrt{2} \,\,\,.
\end{equation}
Eq.~(\ref{eq:rhn}) then yields the effective interactions
\begin{equation}
{\cal L}_{eff} = -\frac{g_2 x}{\sqrt{12}} \left(\frac{M}{M_F}\right)^2 \overline{\nu^c_R} \not\!\!A_{\bf 2_{1/2}}^i
\epsilon_{ij} \ell^j_L - \frac{g_2 x}{\sqrt{3}} \left(\frac{M}{M_F}\right)^2 \overline{\nu^c_R}
\not\!\!A_{\bf 1_{-1}}e_R^c + \mbox{ h.c. }
\end{equation}
Again the decays vanish as $M_F\rightarrow \infty$ or $x\rightarrow 0$.  It is worth mentioning that the 
additional right-handed singlet in the G$_2$ model may provide a possible dark matter candidate, but it is 
clearly premature to pursue that issue in detail here.

As mentioned in the introduction, the exotic gauge boson states in the SU(3) and G$_2$ model can be
long-lived if $M_F$ is sufficiently high.  In the case of the ${\bf 2_{\pm 3/2}}$ gauge bosons,
which is particularly interesting since it contains a doubly-charged bilepton state, we find
\begin{equation}
c \, \tau_X = \mbox{ 0.007 cm } \cdot \left(\frac{\mbox{ 3 TeV}}{M_X}\right)^5 \cdot
\left(\frac{M_F}{\mbox{10 TeV}}\right)^4 \cdot \left[ \frac{g_3^2 (1+x^2)^2}{4 x^2}\right] \,\,\,,
\end{equation}
where we have assumed mixing with one Standard Model generation only and taken $\lambda^e=\lambda^\ell = 
\sqrt{2} m_e/ v$ as a representative choice.  In the three-generation case, the couplings $\lambda^e$
and $\lambda^\ell$ may lead to flavor-violating decays, since we have no restriction on their
values in the basis in which the Standard Model lepton mass matrices are diagonal.  If the lifetime is long,
one could imagine that the branching fraction to lepton flavor-violating decay modes could be substantial
without conflicting with low-energy bounds on lepton flavor-violating processes.  The detailed collider
physics of this possibility seems worthy further investigation.

\section{Conclusions} \label{sec:concl}

In this paper we have considered two extensions of the SU(3) electroweak model~\cite{kd}.  First, 
we introduced new fields (the $\chi$ field in the Higgs sector and the heavy vector-like fermions $\psi$) 
that provide an origin for higher-dimension operators that contribute to exotic gauge boson decays.  We 
also considered a  model that embeds SU(3) in next smallest possible group, G$_2$.  The G$_2$ model includes
exotic gauge bosons that are lighter than those of the SU(3) model by a factor of $\sqrt{2/3}$ and may 
be somewhat easier to produce at future collider experiments.

We analyzed the experimental constraints on these models using precision electroweak observables, following
the approach of Cs\'{a}ki, {\em et al}.~\cite{cekt}.  We noted that measurements of the $W$ mass from LEP II
leads to improved bounds on the original SU(3) electroweak model, but that the effect on the allowed
parameter space is small.   In our models, the dependence on a new parameter $x$, which
encodes a ratio of Higgs field vevs, has a more noticeable effect on the region of parameter space
that is excluded by electroweak constraints.  However, the smallest possible values of the unification scale
and the ${\bf 2_{3/2}}$ gauge boson mass can only be slightly improved using this additional degree of 
freedom.  The constraints on the G$_2$ model were qualitatively similar to those for the SU(3) model.

We have also discussed the higher-dimension operators that contribute to the decay of otherwise 
exotic stable states, focusing specifically on the gauge boson sector.  In the SU(3) model, we showed 
how these operators are generated by explicitly integrating out a sector of vector-like fermions; in 
the G$_2$ model, we construct similar operators directly, and found that a viable model requires that we 
introduce a phenomenologically harmless, singlet fermion in the low-energy spectrum.  In both models, 
otherwise stable gauge bosons can be arbitrarily long-lived, if the cut-off (or vector-like) scale $M_F$ 
is sufficiently high.  

A natural direction for future study is the detailed collider physics of the exotic gauge bosons
in these models.   A collider study of the production and detection of the extra
gauge bosons in the G$_2$ model does not exist and is timely given that they may be within the reach
of the LHC.  The potential for long lifetimes and lepton-flavor violating decays may lead to
unique signatures in TeV-scale collider experiments.

\begin{acknowledgments}
We thank Josh Erlich for many useful discussions and for his comments on the manuscript.  A.R. thanks the 
Monroe Scholars Program at the College of William and Mary for undergraduate research support during the 
Summer of 2007.  We thank the NSF for support under Grant No.~PHY-0456525.
\end{acknowledgments}
\appendix
\section{Generators of $G_2$}

We use the Dynkin diagram as a starting point for analyzing the group G$_2$.
The Dynkin diagram encodes the simple root structure of the group, and provides all the 
information needed to construct its generators.

\begin{figure}
\includegraphics[scale=.6]{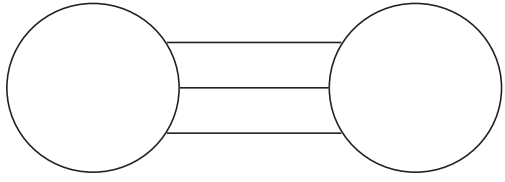}
\caption{Dynkin diagram for group $G_2$.  The two nodes indicate the two simple roots of 
the group.  The triple line connecting them indicates that they are at an angle of 
$\theta=5\pi/6$.\label{fig:dynk}}
\end{figure}

As indicated in Fig.~\ref{fig:dynk}, the group G$_2$ has two simple roots, at a relative
angle of $\theta=5\pi/6$. We may fix the simple roots as
\begin{eqnarray}
\label{eqn:1.1}
\alpha^1&=&\left(0,1\right)\\
\alpha^2&=&\left(\frac{\sqrt{3}}{2},-\frac{3}{2}\right).
\end{eqnarray}
Applying the standard procedure (see Ref.~\cite{georgi}), we obtain the positive roots 
$\{ \alpha^1, \alpha^2, \alpha^1+\alpha^2, 2\alpha^1+\alpha^2, 3\alpha^1+\alpha^2, 
3\alpha^1+2\alpha^2\}$. We also obtain the two fundamental weights
\begin{eqnarray}
\label{eqn:3.4}
\mu^1&=&\left(\frac{\sqrt{3}}{2},\, \frac{1}{2}\right) \\
\mu^2&=&\left(\sqrt{3},\,0\right)
.\end{eqnarray}
The $7$-dimensional, fundamental representation of G$_2$ corresponds to the fundamental weight 
$\mu^1$.  The complete set of weights for this representation are given by 
$\{0, \pm\alpha^1, \pm(\alpha^1+\alpha^2), \pm(2\alpha^1+\alpha^2)\}$. We use this root-weight 
analysis to construct generators for the fundamental representation.  In a convenient basis, the first 
eight generators are given by
\begin{equation}
T^a = \frac{1}{2\sqrt{2}}\left(\begin{array}{ccc}
\lambda_a&0&0 \\
0&-\lambda_a^*&0 \\
0&0&0 \\
\end{array}\right) \,\,\,\,\,a=1\ldots 8 \,\,\,,
\end{equation}
where $\lambda_a$ represents the eight Gell-Mann matrices.  The remaining six generators can
be written
\begin{equation}
T^{9}=\frac{1}{2\sqrt{6}} 
\left(\begin{array}{ccc}0&-i\lambda_2&\sqrt{2}e_3 \\i\lambda_2&0&\sqrt{2}e_3 
\\ \sqrt{2}e_3^T & \sqrt{2}e_3^T&0\end{array}\right)\,\,\,,
\end{equation}
\begin{equation}
T^{10}=\frac{1}{2\sqrt{6}} \left(\begin{array}{ccc}0&-\lambda_2&i\sqrt{2}e_3 
\\-\lambda_2&0&-i\sqrt{2}e_3 \\ -i\sqrt{2}e_3^T & i\sqrt{2}e_3^T&0\end{array}\right)\,\,\,,
\end{equation}
\begin{equation}
T^{11}=\frac{1}{2\sqrt{6}} 
\left(\begin{array}{ccc}0&i\lambda_5&\sqrt{2}e_2 \\-i\lambda_5&0&\sqrt{2}e_2 
\\ \sqrt{2}e_2^T & \sqrt{2}e_2^T&0\end{array}\right)\,\,\,,
\end{equation}
\begin{equation}
T^{12}=\frac{1}{2\sqrt{6}} 
\left(\begin{array}{ccc}0&\lambda_5&i\sqrt{2}e_2 \\ \lambda_5&0&-i\sqrt{2}e_2 
\\ -i\sqrt{2}e_2^T & i\sqrt{2}e_2^T&0\end{array}\right)\,\,\,,
\end{equation}
\begin{equation}
T^{13}=\frac{1}{2\sqrt{6}} 
\left(\begin{array}{ccc}0&-i\lambda_7&\sqrt{2}e_1 \\i\lambda_7&0&\sqrt{2}e_1 
\\ \sqrt{2}e_1^T & \sqrt{2}e_1^T&0\end{array}\right) \,\,\,,
\end{equation}
\begin{equation}
T^{14}=\frac{1}{2\sqrt{6}} \left(\begin{array}{ccc}0&-\lambda_7&i\sqrt{2}e_1 
\\-\lambda_7&0&-i\sqrt{2}e_1 
\\ -i\sqrt{2}e_1^T & i\sqrt{2}e_1^T&0\end{array}\right)\,\,\,,
\end{equation}
where $e_i$ are the unit vectors
\begin{equation}
e_1= \left(\begin{array}{c} 1 \\ 0 \\ 0\end{array}\right),
\,\,\,\,\,
e_2= \left(\begin{array}{c} 0 \\ 1 \\ 0\end{array}\right),
\,\,\,\,\,
e_3= \left(\begin{array}{c} 0 \\ 0 \\ 1\end{array}\right)\,\,\,.
\end{equation}
The SU(3) subgroup relevant to our earlier discussion is formed by the $T^a$ for $a=1\ldots 8$.  In
this basis, the SU(3) decomposition of the fundamental representation of G$_2$ is clear:
\begin{equation}
\mathbf{7}=\left(\begin{array}{c} \mathbf{3} \\ \mathbf{\bar{3}}\\ \mathbf{1}\end{array}\right)
\end{equation}
The generators $T^a$ for $a=9\ldots 14$ can be rewritten as
\begin{equation}
T^a=\frac{1}{2\sqrt{6}} \left(\begin{array}{ccc} 0& M(\chi_a) &\sqrt{2}\chi_a \\ 
M(\chi_a)^\dagger &0 & \sqrt{2}\chi_a^* \\ \sqrt{2}\chi_a^\dagger & \sqrt{2}\chi_a^T&0 \end{array}\right) \,\,\,,
\end{equation}
where $M(\chi_a)$ is a three-by-three matrix defined by
\begin{equation}
M(\chi)^{ij} = - \epsilon^{ijk} \chi^*_k  \,\,\,,
\end{equation}
and where $\chi_9=e_3$, $\chi_{10}=i\,e_3$, $\chi_{11}=e_2$, $\chi_{12}=i\,e_2$, $\chi_{13}=e_1$, and
$\chi_{14}=i\,e_1$.  This form indicates that the decomposition of the $14$-dimensional, adjoint 
representation of  G$_2$ under SU(3) is 
\begin{equation}
\mathbf{14}=\mathbf{8}+\mathbf{3}+\bar{\mathbf{3}}\,\,\,,
\end{equation}
where the $\chi_a$ represent a basis for the 3-dimensional representation, and the ${\bf \bar{3}}$ is related to
the ${\bf 3}$ by complex conjugation.

\end{document}